\documentclass[superscriptaddress,twocolumn,showpacs,prb,floatfix]{revtex4}

\usepackage{graphicx}
\usepackage{amsmath}
\usepackage{amssymb}
\usepackage{amstext}
\usepackage{amsfonts}
\usepackage{amsthm}
\usepackage{latexsym}

\begin{document}

\title{Universality-class dependence of energy distributions in spin glasses}
\author{Helmut G.~Katzgraber}
\author{Mathias K\"orner}
\affiliation{Theoretische Physik, ETH H\"onggerberg,
CH-8093 Z\"urich, Switzerland}
\author{Frauke Liers}
\affiliation{Universit\"at zu K\"oln, Institut f\"ur Informatik, Pohligstrasse
1, 50969, K\"oln, Germany}
\author{Michael J\"unger}
\affiliation{Universit\"at zu K\"oln, Institut f\"ur Informatik, Pohligstrasse
1, 50969, K\"oln, Germany}

\author{A.~K.~Hartmann}
\affiliation{Institut f\"ur Theoretische Physik,
Universit\"at G\"ottingen, Friedrich-Hund-Platz 1, 
37077 G\"{o}ttingen, Germany}
\date{\today}

\begin{abstract}
We study the probability distribution function of the ground-state energies
of the disordered one-dimensional Ising spin chain with power-law 
interactions using a combination of parallel tempering Monte Carlo and branch,
cut, and price algorithms. By tuning the exponent of the power-law
interactions we are able to scan several universality classes. 
Our results suggest that mean-field models have a non-Gaussian limiting
distribution of the ground-state energies, whereas non-mean-field models
have a Gaussian limiting distribution.
We compare the results of the disordered one-dimensional Ising chain to 
results for a disordered two-leg ladder, for which large system sizes can be 
studied, and find a qualitative agreement between the 
disordered one-dimensional Ising chain in the short-range universality class
and the disordered two-leg ladder.
We show that the mean and the standard deviation of the ground-state 
energy distributions scale with a power of the system size.
In the mean-field universality class the skewness does not follow
a power-law behavior and converges to a nonzero constant 
value. The data for the Sherrington-Kirkpatrick model seem to be
acceptably well fitted by a modified Gumbel distribution. Finally, we discuss 
the distribution of the internal energy of the Sherrington-Kirkpatrick model at 
finite temperatures and show that it behaves similar to  
the ground-state energy of the system if the temperature is smaller 
than the critical temperature. 
\end{abstract}

\maketitle

\section{Introduction}
\label{introduction}

Averages of physical quantities and their fluctuations play an important 
role in statistical physics; however the knowledge of the ``average'' 
behavior of a quantity often does not provide sufficient information to 
fully characterize a system, especially if the probability distribution 
of the quantity in question is non-Gaussian, e.g., when it has a nonvanishing
skewness. 
Hallmark examples of such distributions are power-law or exponential 
distributions, which in nature occur in relation to 
earthquakes,\cite{hergarten:02}
magnetic fluctuations,\cite{bramwell:01} 
stock markets,\cite{mantegna:99}
directed polymers in a random medium,\cite{kim:91,halpinhealy:91}
coauthorships in publications, the Internet, 
and other complex networks.\cite{albert:02} 
Many of these systems are characterized by the absence of a characteristic 
length scale such that rare events involving large parts of the system become 
important and strongly influence the average of various quantities.

Recently, there has been a renewed interest in the ground-state energy
distribution $P(E)$ and its limiting form $P_\infty(E)$
of the mean-field Sherrington-Kirkpatrick (SK)
spin-glass model\cite{palassini:03a,andreanov:04,boettcher:05,boettcher:05a} 
and of short-range spin glasses in two and three dimensions.\cite{bouchaud:03}
While studies of the mean-field model have found a non-Gaussian
limiting distribution,\cite{palassini:03a,andreanov:04}
the study of small system sizes of two- and three-dimensional
short-range spin glasses\cite{bouchaud:03} have found a Gaussian limiting
distribution in the thermodynamic limit.
This is supported by the fact
that systems with short-range interactions can be subdivided into
smaller subsystems, coupled weakly
enough to contribute almost independently to the total energy and
leading to a Gaussian distribution via the central-limit
theorem;\cite{bouchaud:03}
however it is important to note that the weak coupling between
the ground-state energies of subsystems below or at an ordering
temperature is not self-evident.

Our goal is to consolidate the different limiting cases of 
short-range and long-range interactions in spin glasses\cite{binder:86}
by studying a disordered one-dimensional Ising spin chain with power-law 
interactions.\cite{bray:86b,fisher:88,leuzzi:99,katzgraber:03,katzgraber:03f} 
The model has the advantage over conventional models in that by tuning the
power-law exponent, several universality classes ranging from
mean-field type behavior to a short-range spin glass can be probed for a large
range of system sizes. 
We show that the presence or absence of {\em mean-field 
behavior}\cite{mezard:87} is reflected in the limiting 
distribution of the ground-state energies. We also study a
two-leg short-range spin ladder, where an exact transfer-matrix algorithm can
be applied, in order to compute the ground-state energy distribution
for large system sizes and to obtain a comparison for the results of the
disordered Ising chain with power-law interactions in the 
short-range phase.
Using a large range of system sizes, our results clearly show
that mean-field spin-glass models have a non-Gaussian
limiting distribution with a finite skewness in the thermodynamic limit,
whereas the limiting distributions for nonmean-field models are Gaussian
(also referred to as ``Normal'').
In addition, we also find that the distribution of the internal energy
of mean-field models is non-Gaussian if the temperature is lower than
the critical temperature.

We do not attempt to make a prediction regarding the exact functional form of
the limiting distribution for an arbitrary spin-glass model.
Bouchaud {\em et al.}\cite{bouchaud:03} have shown for small system sizes 
that typical short-range spin-glass models have a Normal limiting
distribution.
This is not the case for the mean-field model, thus posing the question of
whether the limiting distribution falls into one of the standard
three universality classes for the minimum of uncorrelated
variables:\cite{gumbel:58} Gumbel, Fischer-Tippet-Fr\`echet, and
Weibull distributions. The results
of Bouchaud {\em et al.}\cite{bouchaud:03} cannot determine with certainty
which limiting distribution fits the data best.\cite{bouchaud:97}
Our results suggest that a modified Gumbel
distribution\cite{bramwell:01,portelli:01,palassini:03a}
fits the data for the SK model best, although a detailed probing of the tails
of the energy distributions would be required to make a definite statement if
corrections to the modified Gumbel distribution are required.
For finite values of the power-law
exponent we add a quadratic correction to the modified Gumbel distribution and
show that for finite system sizes the data are well described by this
function. In addition, the quadratic correction (Gaussian) dominates for
increasing system size in the nonmean-field universality class, thus showing
that in the thermodynamic limit a Normal distribution is recovered.

General scaling arguments are presented in Sec.~\ref{sec:stats}.
In Sec.~\ref{ladder} we present results on a one-dimensional two-leg ladder
with short-range interactions in order to illustrate the expected results for
a short range model for very large system sizes. Results on the
one-dimensional Ising spin chain with power-law interactions at zero and
finite temperatures are presented in Sec.~\ref{1dchain}.
We conclude in Sec.~\ref{summary}. The numerical methods
used to compute the ground-state energies \cite{hartmann:01,hartmann:04}
are described in the Appendices.

\section{Statistical Description of Data}
\label{sec:stats}

In general, we expect the ground-state energy of a disordered system to be a 
random variable with mean $\langle E \rangle$, standard deviation 
$\sigma_E$, and skewness $\zeta_E$.\cite{press:95}
In this work we study the size dependence of the aforementioned observables.
In particular, we make the ansatz that the mean ground-state energy of
a (one-dimensional) random system scales as 
\begin{equation}
\langle E \rangle/L =  e_\infty + a L^{-\omega} \; ,
\label{eq:mean}
\end{equation}
where $L$ represents the system size (and number of spins) 
and $\omega$ describes the leading finite-size corrections for the
energy per spin. 
We keep the extra factor of $L$ in Eq.~(\ref{eq:mean}), as well as in the
following definitions in order to be able to compare to the exponent estimates
of Ref.~\onlinecite{palassini:03a}. The standard deviation of the ground-state
energy of a general disordered system can be expected to be determined by 
an exponent $\rho$ via
\begin{equation}
\sigma_E/L =  bL^{-\rho} \; .
\label{eq:stddev}
\end{equation}
The skewness of a distribution of $M$ values $\{E_i\}$ is given by
\begin{equation}
\zeta_E = \frac{1}{M}\sum_{i = 1}^{M} 	\left[ 
				\frac{E_i - \langle E \rangle}{\sigma_E} 
					\right]^3 \; ,
\label{eq:def_skew}
\end{equation}
where $\langle E\rangle$ and $\sigma_E$ are given by 
\begin{equation}
\langle E\rangle = \frac{1}{M} \sum_{i = 1}^{M} E_i
\label{eq:meane}
\end{equation}
and
\begin{equation}
\sigma_E^2 = \frac{1}{M - 1}\sum_{i = 1}^{M} \left(E_i - \langle E\rangle 
\right)^2 \; ,
\label{eq:stddeve}
\end{equation}
respectively. Note that the skewness is a dimensionless quantity.
Following previous results by Ref.~\onlinecite{bouchaud:03} we expect the
skewness to decay as 
\begin{equation}
\zeta_E = c_1 + c_2L^{-\gamma}
\label{eq:skewness}
\end{equation}
with $\gamma > 0$. As we shall see later, $c_1 = 0$ for
the short-range limit of the model.
We also want to test whether the scaled probability distribution
functions $P(\epsilon)$ with $\epsilon = (E -
\langle E\rangle)/\sigma_E$ converge to a limiting form
$P_\infty(\epsilon)$ for $L \rightarrow \infty$. 
If this is the case, then
data for the ground-state energies should be scalable via
\begin{equation}
P(E) = \frac{1}{\sigma_E}P_\infty
\left( 
  \frac{E - \langle E\rangle}{\sigma_E}
\right) \; ,
\label{eq:scaling}
\end{equation}
where $\langle E \rangle$ and $\sigma_E$ are given by Eqs.~(\ref{eq:meane}) and
(\ref{eq:stddeve}), respectively.

\section{Two-leg Spin-glass Ladder}
\label{ladder}

To compare results for the one-dimensional Ising spin chain 
with power-law interactions with a simple benchmark model for which 
large system sizes can be studied, we consider a disordered (short-range)
Ising model on a two-leg ladder (see Fig.~\ref{fig:ladder}). The 
couplings $J_{ij}$ between nearest-neighbor spins are chosen 
from a Normal distribution with zero mean and unit standard deviation.
A system of length $L$ is described by the Hamiltonian
\begin{eqnarray}
  {\mathcal H} &=&      \sum_{l=1}^{L} J_{(l,a),(l,b)} 
                        \, S_{(l,a)} \, S_{(l,b)} \nonumber \\
               & & +    \sum_{l=1}^{L-1} \sum_{i=a,b} 
                        J_{(l,i),(l+1,i)} \, S_{(l,i)} \, S_{(l+1,i)}.
\label{eq:ladderH}
\end{eqnarray}
The first summation in Eq.~(\ref{eq:ladderH}) runs over all rungs $l$, 
while the second summation runs over all exchanges between the rungs, 
and $S_{(l,i)} = \pm 1$ is the value of the (Ising) spin on the $i$th 
leg of the $l$th rung of the ladder. The ground-state energy of the 
system can be efficiently calculated with a transfer-matrix 
algorithm.\cite{morgenstern:80,ozeki:90} 
The transfer-matrix algorithm computes the ground-state energy of
a system of size $L$ in $O(L)$ time so that large systems can be studied. 
The disorder average has to be performed explicitly by repeating the 
algorithm for a number of disorder realizations.

\subsection{Numerical Method: Transfer Matrices}
\label{ladder:method}

\begin{figure}
\includegraphics[width=4.5cm]{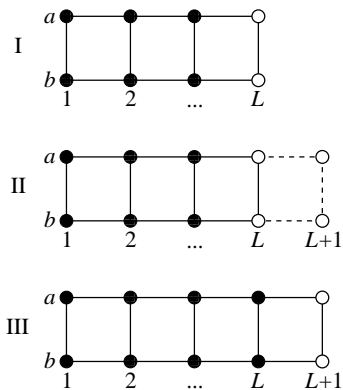}
\caption{
Illustration of a step in the transfer-matrix calculation. Starting with a
system of size $L$ (panel I) whose ground-state energy is known as a function 
of the spins (open circles), we add another rung and calculate the 
change in energy $\Delta_E$ given by the dashed line [see panel II and 
Eq.~(\ref{eq:delta})]. The ground-state energy of the system as a function
of the spins in the $L+1$-th rung is then calculated by taking the minimum of
$E_{\rm g} + \Delta_E$ over all configurations of the $L$th rung 
[see panel III and Eq.~(\ref{eq:minimum})].
}
\label{fig:ladder}
\end{figure}

We can explain the transfer matrix algorithm by starting with a ladder of 
length $L$ and assuming that the ground-state energy 
$E_{\rm g}(L,\{S_{(L,i)}\})$ of the ladder is known as a 
function of the 
spin configuration $\{S_{(L,i)}\}$ of the $L$th rung. 
We add the spins of 
the $L+1$-th rung to the system as illustrated in Fig.~\ref{fig:ladder} 
and use the relation
\begin{eqnarray}
E_{\rm g}(L+1,\{S_{(L+1,i)}\})&& = \min_{\{S_{(L,i)}\}} 
                \left[ E_{\rm g}(L,\{S_{(L,i)}\}) \right. \nonumber \\
        & & +   \left. \Delta_E(\{S_{(L,i)}\},\{S_{(L+1,i)}\}) \right]
\label{eq:minimum}
\end{eqnarray}
to integrate out the spins of the $L$-th rung and to obtain $E_{\rm g}$ as a
function of the spins of the $L+1$-th rung. Here
\begin{eqnarray}
\Delta_E(\{S_{(L,i)}\},\{S_{(L+1,i)}\}) &=& 
                \!\!\! \sum_{i=a,b} \!\!\! J_{(L,i),(L+1,i)} \, 
                S_{(L,i)} \, S_{(L+1,i)} \nonumber \\
                + J_{(L+1,a),(L+1,b)} \!\!\!\! && \!\!\!\!\! 
                S_{(L+1,a)} \, S_{(L+1,b)}
\label{eq:delta}
\end{eqnarray}
is the exchange energy of the spins added on the $L+1$-th rung with themselves
and with the spins of the $L$th rung. Starting with two spins, we iterate 
this procedure until the system has the desired size $L_{\max}$.
The final ground-state energy is then obtained by minimizing over the spins of
the last rung
\begin{equation}
E_g(L_{\max}) = \min_{\{S_{(L_{\max},i)}\}}
\left[ E_{\rm g}(L_{\max},\{S_{(L_{\max},i)}\}) \right]\,.
\end{equation}
We repeat the
calculation until a desired number of disorder realizations is obtained.

\subsection{Results}
\label{ladder:results}

\begin{figure}
\includegraphics[width=\columnwidth]{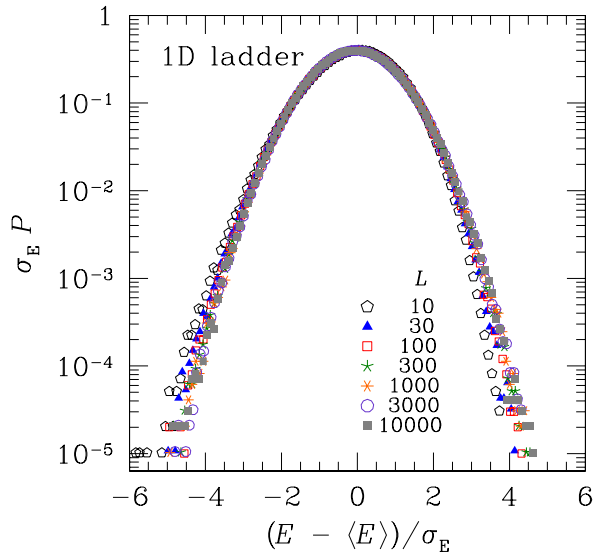}
\caption{(Color online)
Scaling of the ground-state energy according to Eq.~(\ref{eq:scaling}) for 
the ladder system. The data scale well, although deviations in the tails
suggest that the skewness of the function is changing with system size $L$.
}
\label{fig:ladder_scale}
\end{figure}

\begin{figure}
\includegraphics[width=\columnwidth]{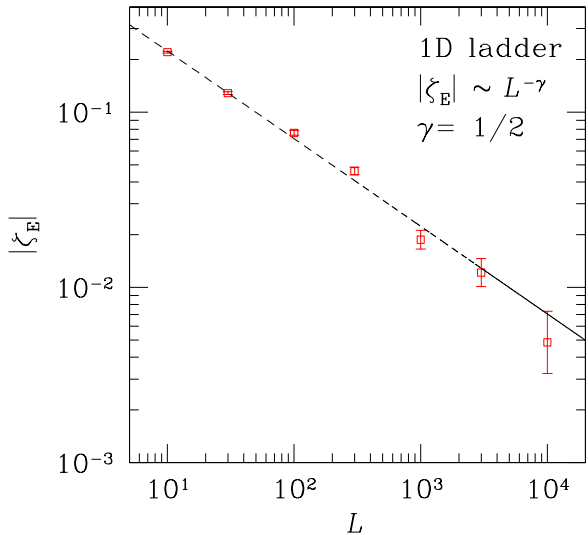}
\caption{(Color online)
Skewness of the energy distributions as a function of system size $L$ for the
ladder system. The skewness can be well fitted to a power-law decay with an
exponent $\gamma \sim 1/2$. This suggests that in the thermodynamic limit the
limiting distribution is Gaussian (zero skewness).
}
\label{fig:ladder_skewness}
\end{figure}

In Fig.~\ref{fig:ladder_scale} we scale the data for the energy of the ladder
system according to Eq.~(\ref{eq:scaling}) for system sizes up to $L = 10^4$.
For each system size we compute $10^6$ samples. The data scale well, although
deviations are present in the tails. In particular, for small $L$ the
distribution is clearly skewed. For the short-ranged ladder system we obtain a
clear power-law decay of the skewness according to Eq.~(\ref{eq:skewness})
with $\gamma \approx 0.5$ (and $c_1 = 0$), 
as can be seen in Fig.~\ref{fig:ladder_skewness}. This suggests that in the
thermodynamic limit the ground-state energies are Gaussian distributed.
For completeness, we quote the results for the size dependence of the mean and
standard deviation. We obtain for the mean energy
\begin{equation}
\langle E \rangle/L = -2.125\,82(8) - 0.801(8)L^{-0.996(4)}
\end{equation}
and thus $\omega \approx 1$. For the fluctuations
\begin{equation}
\sigma_E = 0.976(5)L^{0.497(8)} \; .
\end{equation}
Our results therefore show that $\rho \sim -1/2$ as in the case of the
one-dimensional Ising chain (see below), 
and $\langle E \rangle/L - e_\infty \sim 1/L$.

The scaling of the skewness to zero with a power law and the results 
for very large system sizes already suggest that for short-ranged systems 
the limiting distribution of the ground-state energy is Normal.
This result differs from recent results\cite{palassini:03a} for the mean-field
Sherrington-Kirkpatrick model,\cite{sherrington:75} where the limiting
distribution $P_\infty$ seems to have a finite skewness and thus cannot be
properly described by a Gaussian. Hence, it is desirable to study a
system that 
allows to interpolate between both cases to verify 
whether the change of the distribution coincides with a general change of
the universality class. This is indeed the case for the 
one-dimensional long-range Ising spin glass,
which is studied in the following section.

\section{1D Ising chain}
\label{1dchain}

The Hamiltonian for the one-dimensional long-range Ising spin glass 
with power-law interactions is given by
\begin{equation}
{\cal H} = -\sum_{i,j} J_{ij} S_i S_j \; ,
\label{eq:hamiltonian}
\end{equation}
where $S_i = \pm 1$ represent Ising spins evenly
distributed on a ring of length $L$ in order to ensure periodic boundary 
conditions. The sum is over all spins on the chain and the couplings $J_{ij}$ 
are given by\cite{katzgraber:03}
\begin{equation}
J_{ij} = c(\sigma)\frac{\epsilon_{ij}}{r_{ij}^\sigma}\; ,
\label{eq:bonds}
\end{equation}
where the $\epsilon_{ij}$ are chosen according to a Gaussian distribution
with zero mean and standard deviation unity 
\begin{equation}
{\mathcal P}(\epsilon_{ij}) = \frac{1}{\sqrt{2\pi}}\exp(-\epsilon_{ij}^2/2)
\label{gaussian}
\end{equation}
and $r_{ij} = (L/\pi)\sin[(\pi |i - j|)/L]$ represents the 
{\em geometric} distance between the spins on
the ring.\cite{distances} 
The power-law exponent $\sigma$ determines the range of the
interactions and thus the universality class of the model, as described in the
next section.
The constant $c(\sigma)$ in Eq.~(\ref{eq:bonds}) is chosen to give a mean-field
(MF) transition temperature $T_{\rm c}^{\rm MF} = 1$, where
\begin{equation}
\left(T_{\rm c}^{\rm MF}\right)^2 = \sum_{j\ne i, \; i \; {\rm fixed}} 
[ J_{ij}^2]_{\rm av} = 
c(\sigma)^2 \sum_{j\ne i, \; i \; {\rm fixed}} \frac{1}{r_{ij}^{2\sigma}} \; . 
\label{eq:tcmf}
\end{equation}
Here $[\cdots]_{\rm av}$ denotes an average over disorder.
In this work we compute unscaled energies for the one-dimensional Ising
chain. Thus we find the optimal configuration of spins $\{S_i\}$
that
minimizes the Hamiltonian in Eq.~(\ref{eq:hamiltonian}) for a given set of
interactions $\{J_{ij}\}$, i.e., 
\begin{equation}
E = \min_{\{S_i\}} {\mathcal H}(\{J_{ij}\},\{S_i\}) \, .
\label{eq:energy}
\end{equation}
The (commonly used) energy per degree of freedom $e$ is then given by $e =
E/L$.

\subsection{Phase Diagram}
\label{1dchain:pd}

\begin{figure}
\includegraphics[width=7.0cm]{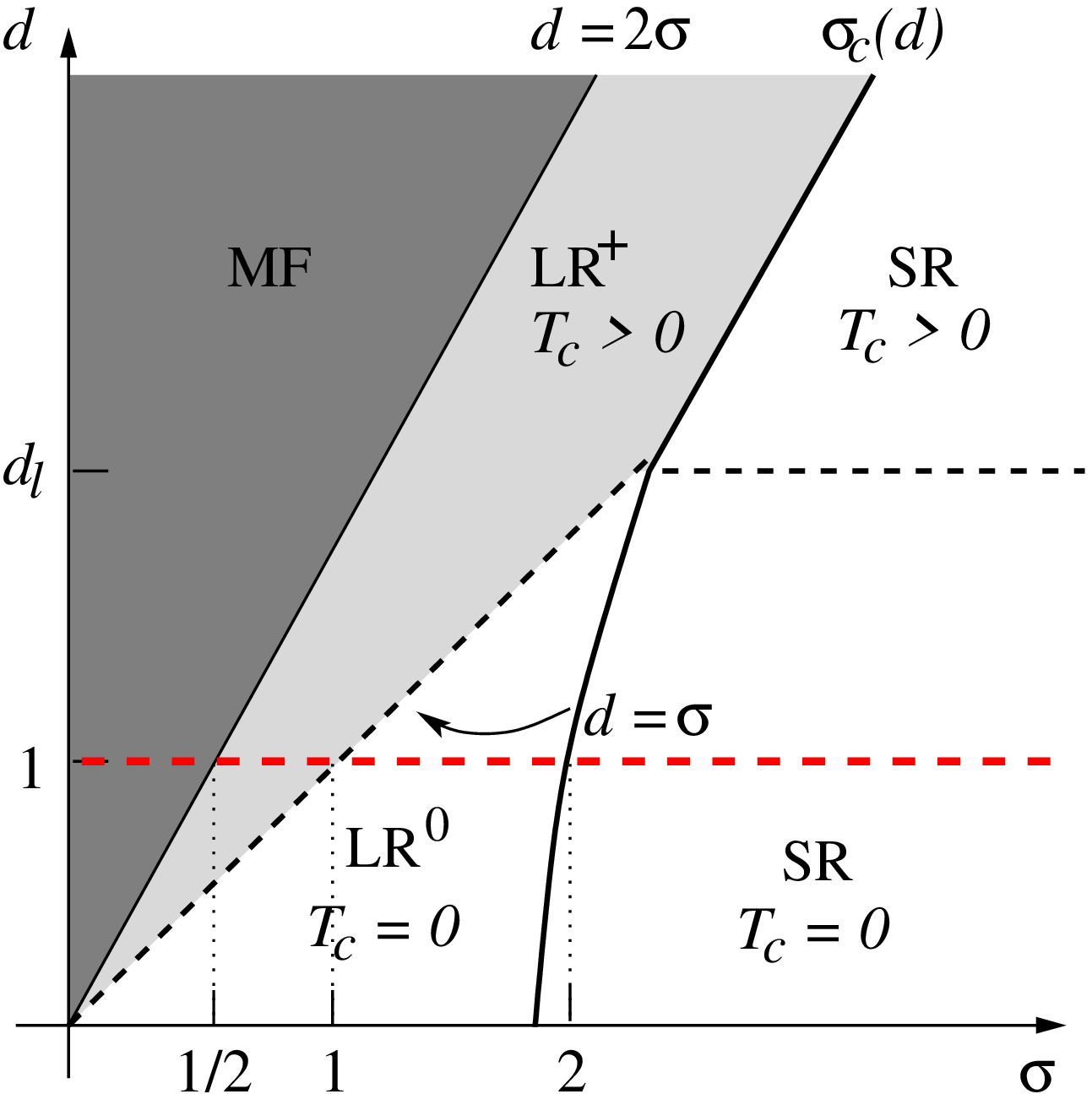}
\caption{(Color online)
Sketch of the phase diagram in the $d$-$\sigma$ plane for the spin-glass
state of the disordered long-range Ising model with power-law interactions
following Ref.~\onlinecite{fisher:88}. The light shaded region (LR$^+$)
is where
there is both a finite $T_{\rm c}$ and the spin-glass state is controlled by the
long-range part of the interaction. 
The thick solid line separates the region of short-range behavior 
(SR) from that 
of long-range behavior and is denoted by
$\sigma_{\rm c}(d)$. The thick dashed line separates
regions where $T_{\rm c} = 0$ (e.g., LR$^0$) from regions where 
$T_{\rm c} > 0$, i.e., it corresponds
to a zero stiffness exponent. The dark shaded region (MF, $\sigma < d/2$) 
is where there is no
thermodynamic limit unless the infinite-range
interactions are scaled appropriately by
the system size. The calculations are performed for $d=1$ (marked by a
horizontal dashed red line), for which $\sigma_{\rm c}(d) = 2$ within a
droplet picture approximation. 
These values of $\sigma$ are marked.
Note that we refer to the
infinite-range region in the phase diagram as ``mean-field
region'' in order to be consistent with previous studies, even
though the mean-field region extends to $d = (2/3)\sigma$.
(Figure adapted from Ref.~\onlinecite{katzgraber:03}.)
}
\label{fig:dsigma}
\end{figure}

The $d$-dimensional long-range Ising spin glass  
with power-law interactions has a very
rich phase diagram in the $d$-$\sigma$ plane. This is summarized in
Fig.~\ref{fig:dsigma}, which is based on work performed by Bray 
{\em et al}.~\cite{bray:86b} and by Fisher and Huse\cite{fisher:88} 
who present a detailed analysis of
the role of long-range interactions within the droplet model. Spin-glass
behavior is controlled by the long-range part of the interaction if
$\sigma$ is sufficiently small, and by the short-range part if $\sigma$ is
sufficiently large. More precisely, one has long-range behavior if the
stiffness exponent\cite{stiffness} of the long-range (LR) universality 
class $\theta_{\rm LR}$ is greater than that of the short-range (SR) 
universality class 
$\theta_{\rm SR}$  and vice versa. In addition, there is an exact result
for $\theta_{\rm LR}$, namely,\cite{bray:86b,fisher:88}
\begin{equation}
\theta_{\rm LR} = d - \sigma ,
\label{eq:theta_lr}
\end{equation}
so long-range behavior occurs if
\begin{equation}
\sigma < \sigma_{\rm c}(d) = d - \theta_{\rm SR}(d)\; .
\label{eq:sigmac}
\end{equation}
Equation (\ref{eq:theta_lr}) indicates that critical exponents depend
continuously on $\sigma$ in the long-range region, even though they are
independent of $\sigma$ in the region controlled by the short-range part
of the interaction. Thus we expect to be able to tune the different
universality classes by changing the exponent $\sigma$.
The condition for a finite-temperature transition is $\theta > 0$, where
$\theta$ refers here to the greater of $\theta_{\rm SR}$ and $\theta_{\rm
LR}$. For the short-range model, there is a finite-temperature transition
(i.e., $\theta_{\rm SR} > 0$) for $d$ larger than the lower critical
dimension $d_l$, which is found numerically to lie between 2 and
3.\cite{bray:84,mcmillan:84,mcmillan:84b,kawashima:96,hartmann:99,%
ballesteros:00,hartmann:01a}
For $d = 1$, as in the present study, we obtain a finite transition
temperature for $\sigma < 1$.
For $\sigma < d/2$, the model would
not have a thermodynamic limit ($T_{\rm c}$
would diverge) if the interactions were not
scaled as shown in Eq.~(\ref{eq:tcmf}). The scaling leads 
to a power-law dependence on $L$ with
a negative exponent, i.e., $c(\sigma) \to 0 $ for $L \to \infty$.
$\sigma = 0$ corresponds to the SK model and 
leads to $c(0)\sim 1/L$.

\subsection{Numerical Methods}
\label{1dchain:numerics}

Ground-state energies for the one-dimensional Ising chain are computed using
the parallel tempering Monte Carlo
method\cite{hukushima:96,marinari:96,moreno:03,katzgraber:03} when the
power-law exponent $\sigma$ is small, and the branch, cut, and price
(BCP) algorithm \cite{barahona:88,liers:04,juenger:95}
when $\sigma$ is large. As reported in
Ref.~\onlinecite{katzgraber:03f}, the time to compute a ground-state instance
using the parallel tempering Monte Carlo method scales in
practice with a power of the
system size for $\sigma \lesssim 1.25$, whereas for large values of $\sigma$
the time to compute a ground state scales $\sim \exp(aL)$, with $a$ a constant.
In this case we use the BCP algorithm which performs
best for short-range interactions, thus ideally complementing the parallel
tempering method. Details about the algorithms used and simulation parameters
can be found in the Appendices. 

\subsection{Results}
\label{1dchain:results}

\begin{figure}
\includegraphics[width=\columnwidth]{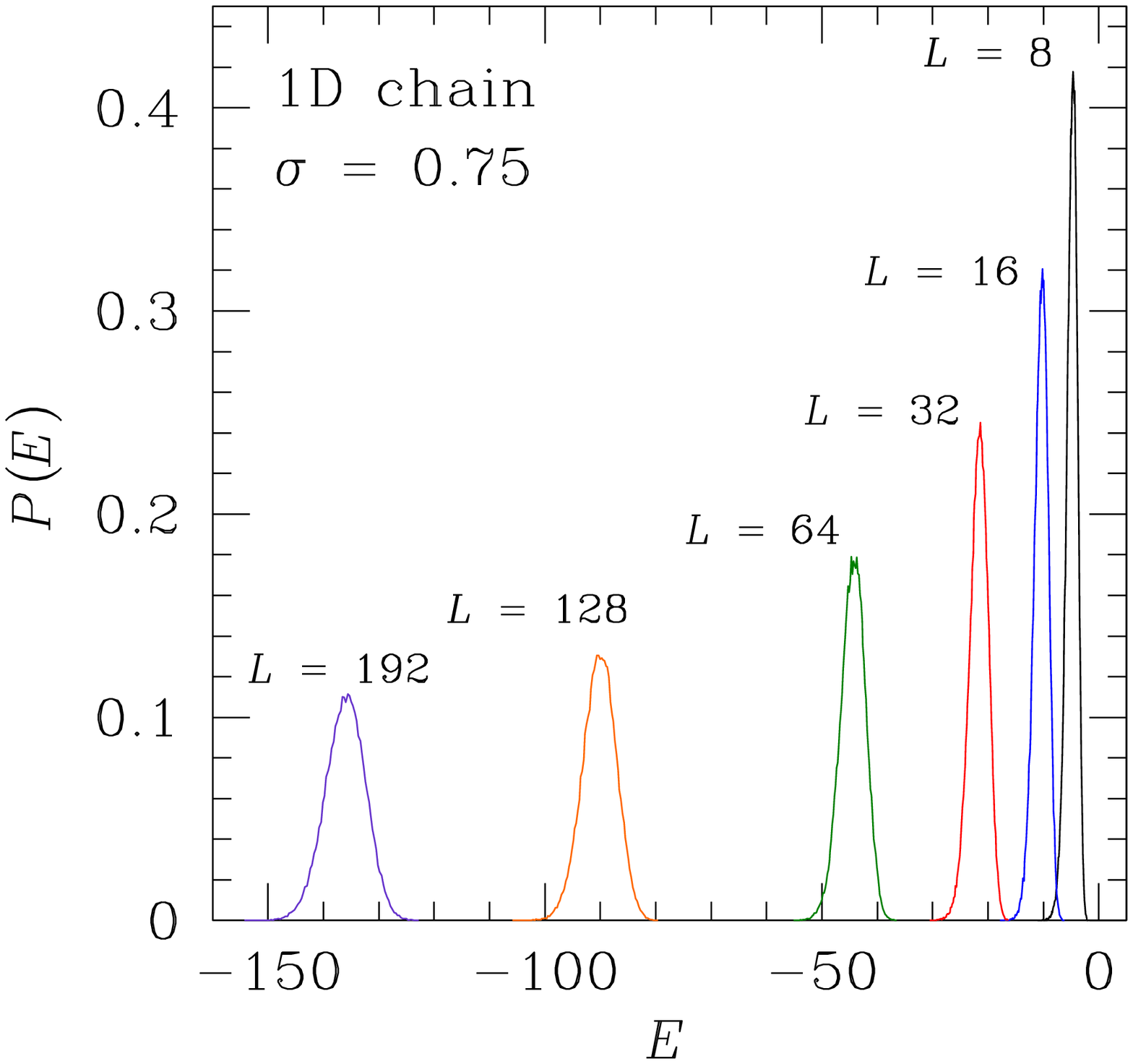}
\vspace*{-1.0cm}
\caption{(Color online)
Unscaled energy distributions for several system sizes for the one-dimensional
Ising chain with $\sigma = 0.75$ (LR$^+$ phase). 
}
\label{fig:e.histo-0.75}
\end{figure}

\begin{figure}
\includegraphics[width=\columnwidth]{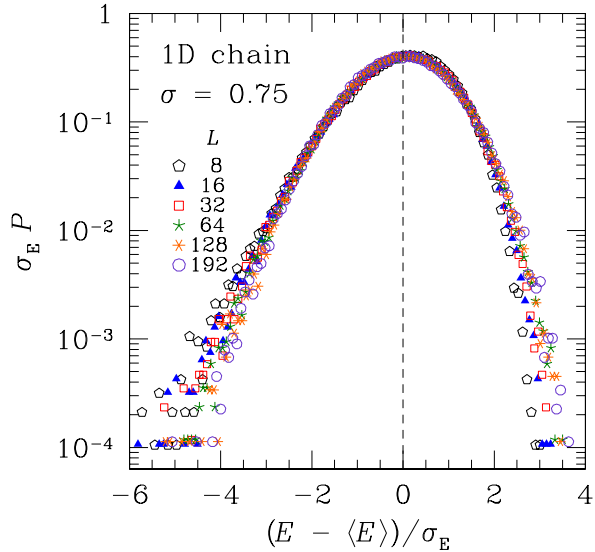}
\vspace*{-1.0cm}
\caption{(Color online)
Scaled ground-state energy distributions for several system sizes for the
one-dimensional Ising chain with $\sigma = 0.75$ (LR$^+$ phase). The dashed
vertical line is a guide to the eye to illustrate the skewness of the
distribution. The spread of the data in the tails suggests that the skewness
changes with system size.
}
\label{fig:s.histo-0.75}
\end{figure}

\begin{figure}
\includegraphics[width=\columnwidth]{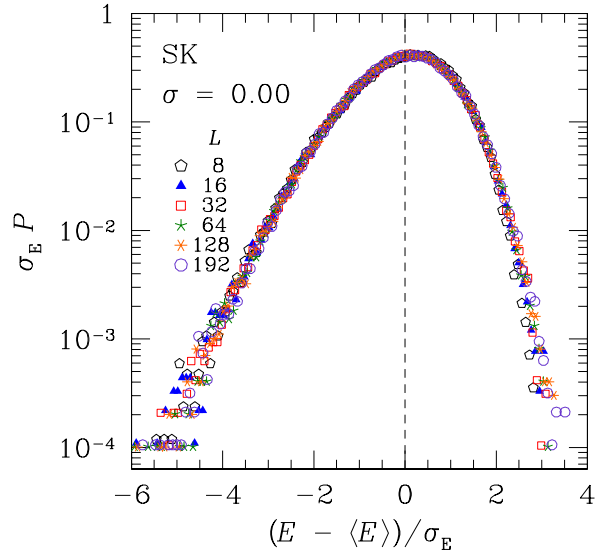}
\vspace*{-1.0cm}
\caption{(Color online)
Scaled ground-state energy distributions for several system sizes for the
one-dimensional Ising chain with $\sigma = 0$ (MF phase, SK model). 
The dashed vertical line is a guide to the eye to illustrate the skewness of 
the distribution. The data show little spread in the tails suggesting a weaker
dependence on $L$ than for larger values of $\sigma$.
}
\label{fig:s.histo-0.00}
\end{figure}

\begin{figure}
\includegraphics[width=\columnwidth]{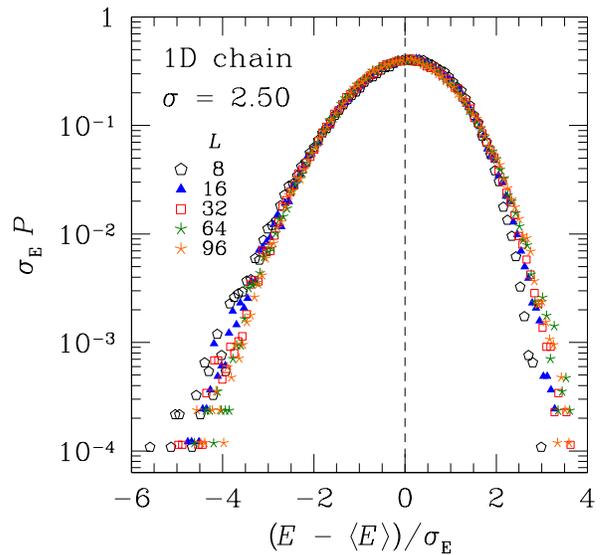}
\vspace*{-1.0cm}
\caption{(Color online)
Scaled ground-state energy distributions for several system sizes for the
one-dimensional Ising chain with $\sigma = 2.50$ (SR phase). The dashed
vertical line is a guide to the eye to illustrate the skewness of the
distribution. The data show a moderate dependence on the system size.
}
\label{fig:s.histo-2.50}
\end{figure}

For each system size we compute $10^5$ ground-state realizations for system
sizes up to $L = 192$ (see Table \ref{tab:numbers} for details).
In Fig.~\ref{fig:e.histo-0.75} we show a representative set of the unscaled
data for $\sigma = 0.75$ in the LR$^+$-phase for several system sizes. Data
for other values of $\sigma$ show a similar qualitative behavior.
The data in Fig.~\ref{fig:e.histo-0.75} can be scaled according to
Eq.~(\ref{eq:scaling}), the result is displayed in
Fig.~\ref{fig:s.histo-0.75}.
The data are clearly skewed and the tails indicate that the skewness depends
on the system size. In Fig.~\ref{fig:s.histo-0.00} we show scaled data for the
ground-state energy distributions for $\sigma = 0$ (SK limit, MF phase).
The data also show a clear asymmetry, but the spread in the tails is
noticeably smaller than for larger values of $\sigma$ (see 
Figs.~\ref{fig:s.histo-0.75} and \ref{fig:s.histo-2.50}) suggesting a 
smaller dependence of the skewness of the distribution on the system size.

\begin{figure}
\includegraphics[width=\columnwidth]{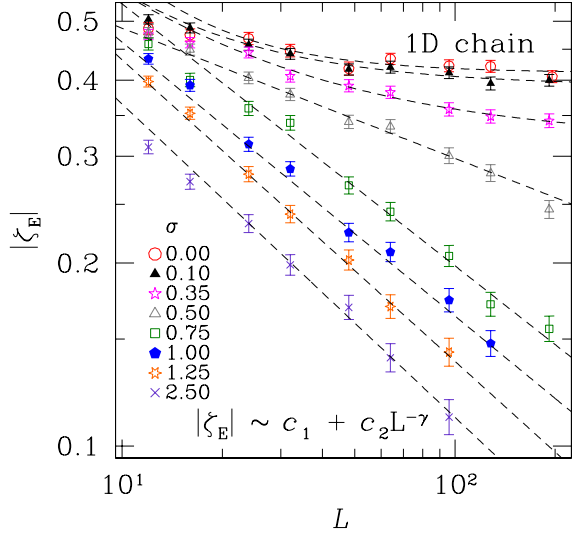}
\vspace*{-1.0cm}
\caption{(Color online)
Skewness $\zeta_E$ as a function of system size $L$ for several values of
$\sigma$. For $\sigma < 0.5$ (MF phase) the data scale as $\zeta_E 
\sim c_1 + c_2 L^{-\gamma}$ with $c_1 > 0$ [Eq.~(\ref{eq:skewness})] 
thus tending to a constant in the thermodynamic limit. 
For $\sigma \ge 0.5$ the skewness decays with a power-law behavior, i.e., 
$c_1 = 0$ (fits done for $L \ge 64$). Note that for $\sigma > 1$, for
which $T_{\rm c} = 0$, $\gamma \approx 0.5$ in agreement with the results for
the ladder system presented in Sec.~\ref{ladder}.
}
\label{fig:skewness}
\end{figure}

In order to better quantify the aforementioned behavior, in
Fig.~\ref{fig:skewness} we present data for the skewness as a function of
system size for several values of the power-law exponent $\sigma$.
The data show that for $\sigma > 0.5$ the skewness of the ground-state energy
distributions decays with a power law 
$|\zeta_{\rm E}|\sim L^{-\gamma}$, with 
$\gamma \approx 0.5$ in the SR phase, whereas for $\sigma \leq 0.5$ 
(MF phase) the
skewness is well fitted by Eq.~(\ref{eq:skewness}) with $c_1 > 0$ thus tending
to a constant in the thermodynamic limit. This means that the mean-field
models present a singular behavior in which the ground-state energy
fluctuations are non-Gaussian in the thermodynamic limit. This is not the case
for the nonmean-field universality class where a limiting Gaussian 
behavior is obtained for $L
\rightarrow \infty$. Note that $\gamma \approx 0.5$ for $\sigma > 1$ for
which $T_{\rm c} = 0$, in agreement with the results for the ladder system
studied in Sec.~\ref{ladder}.

\begin{figure}
\includegraphics[width=\columnwidth]{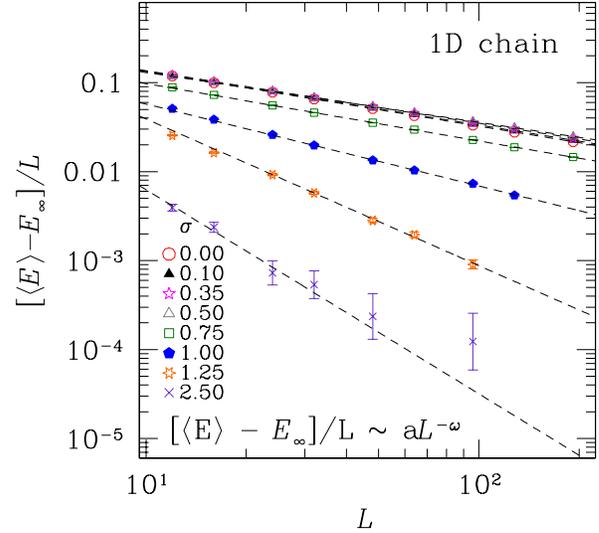}
\vspace*{-1.0cm}
\caption{(Color online)
Mean $[\langle E \rangle - E_\infty]/L$ as a function of system size 
$L$ for several values of $\sigma$. The data are expected to decay as a power
of the system size with an exponent $\omega$. Note that for the SK limit
$\omega \approx 2/3$ at $T = 0$ in agreement with other
predictions (Refs.~\onlinecite{boettcher:03}, \onlinecite{palassini:03a},
\onlinecite{bouchaud:03}, and \onlinecite{andreanov:04})
[see Fig.~\ref{fig:omega} for $\omega(\sigma)$] and that for 
$\sigma=2.5$ we obtain $\omega = 2.29(17)$.
}
\label{fig:mean}
\end{figure}

\begin{figure}
\includegraphics[width=\columnwidth]{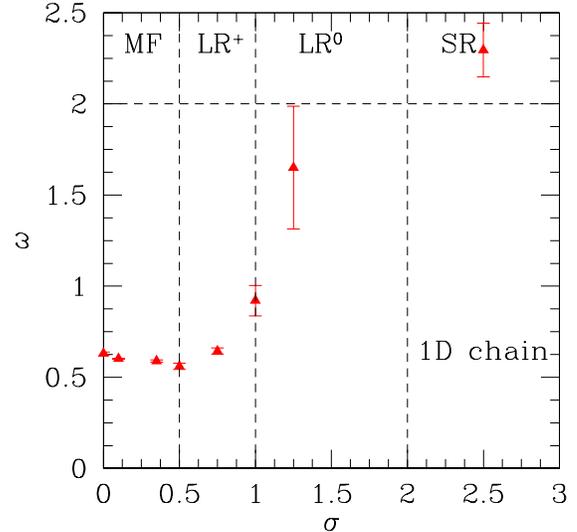}
\vspace*{-1.0cm}
\caption{(Color online)
Exponent of the mean energy ($\omega$) as a function of $\sigma$, according to
Eq.~(\ref{eq:mean}). $\omega$ increases from the SK value ($\sim 2/3$) for
increasing $\sigma$.
The exponents are only estimated for the four largest system sizes
studied for a given value of $\sigma$. See Table \ref{tab:numbers} for
details. In this and following figures, the boundaries between the different
universality classes are denoted by vertical dashed lines.
}
\label{fig:omega}
\end{figure}

We also study the size-dependence of the mean energy as a function of
$\sigma$. For the mean-field Sherrington-Kirkpatrick 
model\cite{sherrington:75} ($\sigma = 0$) it is 
known that 
$\omega \sim 2/3$.\cite{boettcher:03,palassini:03a,bouchaud:03,andreanov:04}
Our results agree well with this prediction, i.e., $\omega = 0.64(1)$ (the
quality of fit probability\cite{press:95} is $Q = 0.51$; the fit is performed
for $L \ge 64$). Unfortunately, there are no predictions for the different
exponents for $\sigma > 0$, thus we will focus on comparing 
the present results to data for the SK model.
In Fig.~\ref{fig:mean} we show data for all values of $\sigma$ studied. For
increasing $\sigma$, $\omega$ increases rapidly and then saturates at $\omega
\approx 2$ in the SR phase. This can be understood by studying the model for
$\sigma \rightarrow \infty$. In this limit there is no frustration, except
that with a 50 \% probability there will be a broken bond due to the periodic
boundary conditions. Since the weakest bond will be broken, for a
continuous distribution the energy scales as $\sim 1/L$. Since the total
energy scales with system size, we expect the finite-size correction to the 
average energy per spin to be $\sim 1/L^2$, i.e., $\omega = 2$.
This behavior can be seen in 
Fig.~\ref{fig:omega} where
we show the behavior of $\omega$ [see Eq.~(\ref{eq:mean})] in detail for 
the different universality classes.

\begin{figure}
\includegraphics[width=\columnwidth]{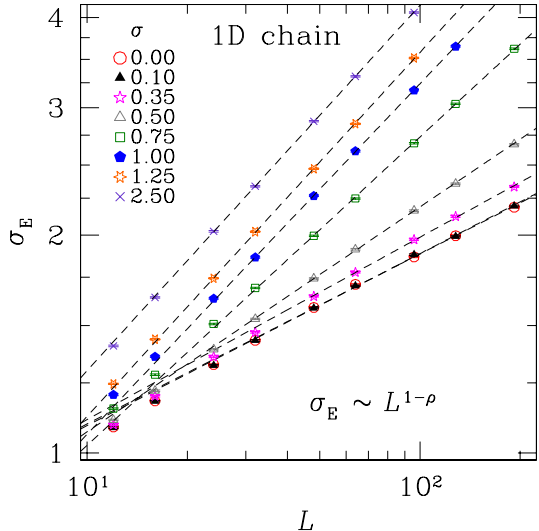}
\vspace*{-1.0cm}
\caption{(Color online)
Standard deviation $\sigma_E$ as a function of system size
$L$ for several values of $\sigma$. The data for $\sigma_E/L$ are expected 
to decay as a power of the system size with an exponent $\rho$. 
Note that for the SK limit
$\rho \approx 3/4$, in agreement with other 
predictions (Ref.~\onlinecite{aspelmeier:02}, \onlinecite{palassini:03a},
and \onlinecite{bouchaud:03})
[See Fig.~\ref{fig:rho} for $\rho(\sigma)$].
}
\label{fig:stddev}
\end{figure}

\begin{figure}
\includegraphics[width=\columnwidth]{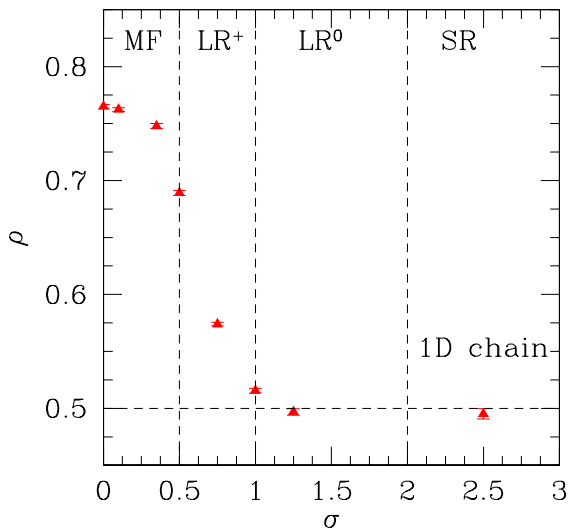}
\vspace*{-1.0cm}
\caption{(Color online)
Exponent for the energy fluctuations $\rho$ as a function of $\sigma$
[see Eq.~(\ref{eq:stddev})]. For $\sigma \rightarrow 0$ $\rho \approx 3/4$ in
agreement with Refs.~\onlinecite{aspelmeier:02}, \onlinecite{bouchaud:03}, 
and \onlinecite{palassini:03a}. For  $\sigma \rightarrow \infty$, $\rho
\rightarrow 1/2$, as predicted by the central limit theorem, and in agreement
with the results on the ladder system presented in Sec.~\ref{ladder:results}.
}
\label{fig:rho}
\end{figure}

The behavior of the energy fluctuations is shown in Fig.~\ref{fig:stddev} as a
function of system size $L$ for several values of $\sigma$. In the SK limit
there are contradicting predictions regarding the power-law exponent $\rho$ 
of the energy fluctuations $\sigma_E$. While 
Crisanti {\em et al.}\cite{crisanti:92}
find $\rho = 5/6$, Bouchaud {\em et al.}\cite{bouchaud:03}
and Aspelmeier {\em et al.}\cite{aspelmeier:02} find $\rho = 3/4$. In this
work we obtain $\rho = 0.775(2)$ ($Q = 0.58$; fits done for $L \ge 64$), which
is also in agreement with the work by Palassini.\cite{palassini:03a}
In Fig.~\ref{fig:rho} we show the $\sigma$ dependence of $\rho$. It is
noteworthy that $\rho$ decreases from the mean-field value $\sim 3/4$ to $1/2$
in the short-range universality class. This is to be expected as for $\sigma
\rightarrow \infty$ the central limit theorem predicts that $\rho =
1/2$.\cite{bouchaud:03} 
Note that the results found agree with the prediction of the short-range
ladder system in Sec.~\ref{ladder:results}.

\subsection{Limiting Distribution}
\label{1dchain:area}

\begin{figure}
\includegraphics[width=\columnwidth]{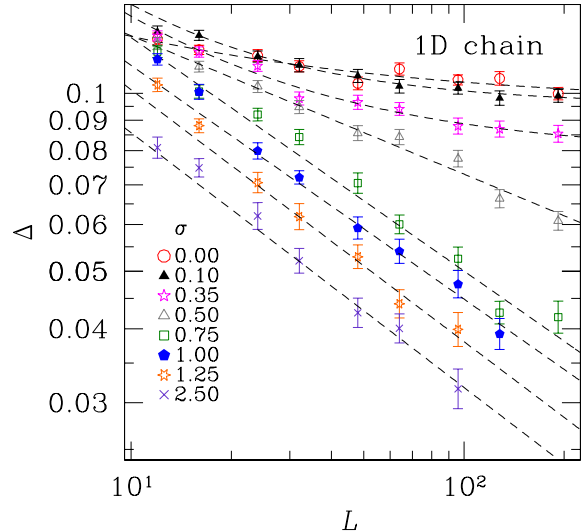}
\vspace*{-1.0cm}
\caption{(Color online)
Difference in area between the actual data for the energy probability
distributions of the one-dimensional Ising chain to a Gaussian limiting
distribution as a function of system size $L$ for several values of $\sigma$
[see Eq.~(\ref{eq:Delta})]. In the MF phase ($\sigma < 0.5$) the area
difference tends to a constant for increasing system size, whereas for $\sigma
\ge 0.5$ the area difference decays with a power of the system size. Note the
close resemblance to the behavior found for the skewness of the distribution,
Fig.~\ref{fig:skewness}.
}
\label{fig:Delta}
\end{figure}

In order to further strengthen the conjecture that ground-state energy
distributions remain skewed in the thermodynamic limit for the mean-field
phase, in this section we study the area deviation of the 
normalized energy distributions
in comparison to a Normal distribution $N(\epsilon)$. We define
the area difference $\Delta$ via
\begin{equation}
\Delta = \int_\epsilon |P(\epsilon) - N(\epsilon)| d\epsilon \; ,
\label{eq:Delta}
\end{equation}
where $P(\epsilon)$ are the actual rescaled data [Eq.~(\ref{eq:scaling})]. 
In Fig.~\ref{fig:Delta} we show the
area difference 
as a function of system size $L$ for several values of $\sigma$. The data for 
$\sigma < 0.5$ can be well fitted by a functional form $\sim f +g/L^h$, i.e., 
the area difference tends to a nonzero constant in the thermodynamic
limit. This is not the case for $\sigma \ge 0.5$ where the area difference
decays with a 
power law of the system size, thus showing that the difference between the
data and a Gaussian limiting distribution decreases for increasing $L$.

\begin{figure}
\includegraphics[width=\columnwidth]{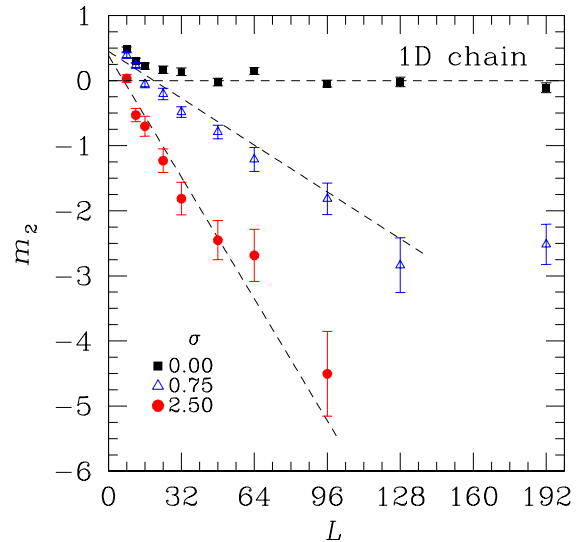}
\vspace*{-1.0cm}
\caption{(Color online)
Coefficient $m_2$ to the quadratic correction
term in the modified Gumbel distribution, Eq.~(\ref{eq:gumbel}), 
as a function of system size for several
values of the power-law exponent $\sigma$. The data show that $m_2$
converges to a value close to zero
for the SK model in the thermodynamic limit thus suggesting that the energy
distributions of the SK model are possibly 
well described by a modified Gumbel distribution function
in agreement with results from Ref.~\onlinecite{palassini:03a}. 
For all
$\sigma > 0$, $m_2$ tends to a finite negative value in the thermodynamic 
limit. For large values of $\sigma$, $m_2$ dominates thus showing that in the SR
universality class the limiting probability distribution is well described by
a Gaussian. The dashed lines are guides to the eye.
}
\label{fig:m2}
\end{figure}

Palassini\cite{palassini:03a} has fitted the data for the scaled probability 
distribution functions of the SK model, Eq.~(\ref{eq:scaling}), to a 
modified Gumbel distribution\cite{gumbel:58,bramwell:01,portelli:01} 
$g_{m}(\epsilon)$, and finds
good agreement between the data and the fit, especially when studying the 
cumulative distributions $Q(\epsilon) = \int^\epsilon P(x) dx$. 
In addition, Palassini shows
that the best fit seems to be obtained for $m = 6$, although to date it is
unclear why the aforementioned value of $m$ fits the data best. Because 
outside the MF universality class the limiting distribution function seems to 
converge to a Normal distribution, we modify the standard modified Gumbel
distribution by taking into account a Normal contribution,\cite{hartmann:02e}
i.e., 
\begin{equation}
g^\prime_{m}(\epsilon) = N(y) g_{m}(y)\; ,
\label{eq:gumbel}
\end{equation}
where
\begin{equation}
y = \frac{\epsilon - \mu}{\nu} \;
\label{eq:gumbel2}
\end{equation}
and 
\begin{equation}
g_{m}(\epsilon) = w_1 \exp \left[ m y - m e^y \right] 
\label{eq:mgumbel}
\end{equation}
is the modified Gumbel distribution and
\begin{equation}
N(\epsilon) = w_2 \exp \left[ m_2 y^2 \right]
\label{eq:normal}
\end{equation}
is a Normal distribution.  Here $\mu$ is the most-probable value,
$\nu$ a standard deviation, and $w_i$ represents an overall
normalization factor. For simplicity, we can fix $m = 6$ and study the 
behavior of the coefficient $m_2$ as a function of system size for different 
values of $\sigma$. Note that for $m_2 = 0$ $g^\prime_m(x) \propto g_m(x)$, up
to a global scaling factor. A multiplicative ansatz [instead of, for
example, an additive ansatz of the form $aN(\epsilon) +
bg_m(\epsilon)$] can be motivated by keeping in mind that for
short-range interactions, the system can be divided into subsystems
which contribute almost independently to the total energy.  In
Fig.~\ref{fig:m2} we show data for $m_2$ versus $L$ for a few
representative values of $\sigma$. Our results show that for $\sigma =
0$ (SK model) $m_2$ converges to a value close to zero
for $L \rightarrow \infty$. For
$0 < \sigma \le 0.5$ the limiting distribution is non-Gaussian, yet
$m_2$ is small, but finite. For $\sigma > 0.5$ the Gaussian
contribution via $m_2$ dominates in the thermodynamic limit (at least for a 
finite fitting region), as can be
seen in Fig.~\ref{fig:m2}.
This shows that the energy distributions in the SK model can be
well described in the thermodynamic limit by a modified Gumbel distribution. 
In order to test the existence of small Gaussian corrections to the modified
Gumbel distribution for the SK model, large-scale simulations probing the tails
of the distribution function in detail would be required which are beyond the
scope of this work.
For all other values of $\sigma < 0.5$ there are clearly Gaussian 
corrections to the Gumbel distribution, whereas for $\sigma \ge 0.5$ the data 
in the thermodynamic limit are well described by a Normal limiting 
distribution. This could be due to the fact that there are no length scales
associated with the mean-field model. Thus any length-scale associated effects
will scale with system size. This is not the case in the short-range models
where a length scale will not necessarily scale with system size, therefore
yielding a Normal distribution in the thermodynamic limit.

\subsection{Finite Temperatures}
\label{1dchain:tgt0}

We want to test if the fact that the ground-state energy distribution of the
SK model is skewed in the thermodynamic limit is a unique property of the
ground state, or if similar effects can be observed at finite temperatures.
Because the parallel tempering Monte Carlo method used to compute the
ground-state energies of the one-dimensional Ising chain at small values of
$\sigma$ requires the system to be simulated at several temperatures ranging
to values well above the spin-glass transition (for the SK model 
$T_{\rm c} = 1$), we have also studied the behavior of the internal energy 
distributions in the mean-field limit as a function of temperature.
The internal energy $U$ for a given disorder realization $\{J_{ij}\}$ is given
by 
\begin{equation}
U = \left<{\mathcal H}(\{J_{ij}\},\{S_i\}) \right> \; ,
\label{eq:u}
\end{equation}
where the Hamiltonian ${\mathcal H}$ is given by Eq.~(\ref{eq:hamiltonian}).
Here $\langle \cdots \rangle$ represents a thermal average over $t_{\rm eq}$
Monte Carlo steps that
we perform after having equilibrated the system for 
a time $t_{\rm eq}$ (see Table \ref{tab:numbers} for details).

\begin{figure}
\includegraphics[width=\columnwidth]{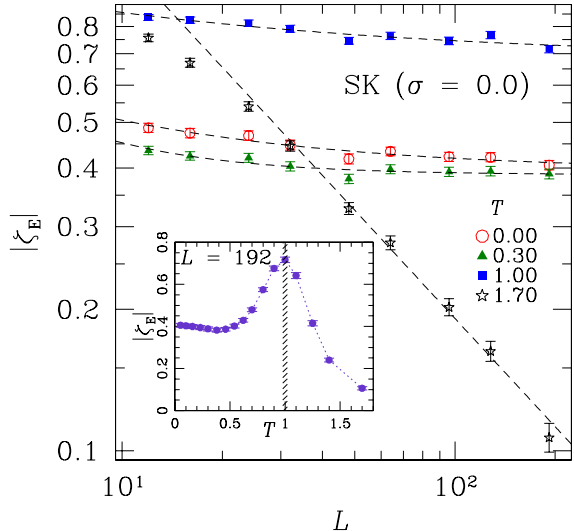}
\vspace*{-1.0cm}
\caption{(Color online)
Skewness of the internal energy probability distribution functions of the SK
model as a function of system size for different temperatures. The data show a
curvature for $T \le T_{\rm c}$ in a log-log scale thus suggesting that the
skewness converges to a constant value in the thermodynamic limit. For $T >
T_{\rm c} = 1$ the skewness decays with a power of the system size (Normal
limiting probability distribution function). The inset shows the skewness of
the internal energy distribution of the SK model for $L = 192$ (largest system
size studied) as a function of temperature. The data show that for finite
system sizes the skewness seems to peak at the transition ($T_{\rm c} = 1$,
shaded area).
}
\label{fig:finite-T}
\end{figure}

Figure \ref{fig:finite-T}
shows data for the skewness of the internal energy distributions as a function
of system size for several temperatures ranging from the ground-state to well
above the critical temperature. The results show that the skewness of the
distributions tend to a constant value in the thermodynamic limit for $T \le
T_{\rm c}$ 
(curved fitting functions in a log-log plot, Fig.~\ref{fig:finite-T}),
thus showing that skewed energy distributions seem to persist for
any temperature below the critical temperature. For temperatures above the 
critical temperature, the skewness shows again a power-law behavior
thus suggesting that for $T > T_{\rm c}$ the limiting distribution is Normal,
as one would expect. Therefore, the limiting probability distribution
is skewed in the thermodynamic limit for all temperatures below the 
critical point. 

The inset of Fig.~\ref{fig:finite-T} shows the skewness of the probability
distribution function of the internal energy of the SK model for $L = 192$
as a function of temperature. The data show a peak around $T_{\rm c} = 1$. 
We expect the functional
form of the ground-state energy distribution
to remain approximatively the same for $L \rightarrow
\infty$ when $T < T_{\rm c}$, whereas for $T > T_{\rm c}$ we expect
for the skewness $\zeta_{\rm E} \rightarrow 0$ in the thermodynamic limit.
It would be interesting to understand the origins of this behavior of
the mean-field model analytically.

For nonzero values of $\sigma$ we find finite-temperate results in agreement 
with the data presented in Sec.~\ref{1dchain:results}: The distributions 
become Normal in the thermodynamic limit.

\section{Summary and Conclusions}
\label{summary}

We have studied in detail the probability distribution function of the
ground-state energy of the one-dimensional Ising spin chain with random
power-law interactions for several values of the power-law exponent $\sigma$.
Using sophisticated parallel-tempering methods 
(fast for small values of $\sigma$) and
a branch, cut, and price algorithm (fast for large values of $\sigma$),
relatively large
system sizes have been studied over the full range of the parameter $\sigma$.

For the SK limit, when $\sigma = 0$, our results agree with previous numerical
work by Palassini.\cite{palassini:03a} We find by studying different
moments of the distribution, that the SK model has a skewed probability 
distribution function in the thermodynamic limit that 
is well fitted by a modified Gumbel distribution, possibly with small Gaussian
corrections. This
behavior is not only valid for the ground-state energy, but also for
energies below the critical temperature.

By varying the power-law exponent $\sigma$ we scan several universality
classes and show that for the nonmean-field regime when $\sigma > 0.5$ the 
probability distribution functions converge to a Normal distribution in the 
thermodynamic limit, in agreement with a short-range spin-glass ladder. 
Thus a skewed ground-state energy probability distribution function is a 
characteristic property of the mean-field spin-glass model and the
change of the distribution's characteristic coincides with the transition
line between the MF and LR universality classes.
This behavior
again poses the question, of whether
the mean-field description of low-temperature properties of spin glasses is
adequate for nonmean-field models, as has been observed previously
by studying other measurable 
quantities,\cite{palassini:99,krzakala:00,katzgraber:01,young:04,newman:03}
although other studies\cite{marinari:00a} have found different results.

Thus far it is unclear to us why the limiting distribution for the
SK case is well described by a modified Gumbel distribution with parameter 
$m > 1$, i.e., an extreme-value distribution for selecting the $m$th 
smallest value out 
of a large number of $M$ uncorrelated values.\cite{gumbel:58}
If all $2^N$ energy levels of a
system with $N$ spin were uncorrelated, then the ground state would be simply
the minimum of all $2^N$ uncorrelated values and a standard Gumbel
distribution ($m=1$) would be the limiting distribution. Clearly the energy 
values of a spin glass are not fully uncorrelated, but
recently it has been observed\cite{bauke:04} 
that the energy levels of the Edwards-Anderson model behave at least 
locally (i.e., in small intervals) like
a random-energy model. This might be the underlying reason why a Gumbel
distribution seems to describe the data best, as well as for the 
occurrence of a nonvanishing skewness in the MF case for $\sigma < 0.5$.

In general, we see that by studying the distributions of measurable
quantities such as for the ground-state energy, we have another approach to
discriminate mean-field-type behavior from simpler structures of the phase
space. Therefore this approach supplements other numerical means of studying
the organization of phase space, such as calculating the distributions of
overlaps,\cite{bhatt:85} clustering configurations,\cite{barthel:04} 
or the calculation of correlation-matrix eigenvalues.\cite{sinova:00}
Hence, it should be fruitful to study the distributions of ground-state 
energies in detail also for other models. This is especially interesting 
when a disorder-driven phase transition occurs, such as for parametrized 
random bond models, random-field systems, or optimization algorithms on 
random graphs. So far the  body of the ground-state energy distributions has
been tested in detail. More information about the tails of the 
distributions could be accessed using rare-event 
techniques\cite{hartmann:02e} also for the standard spin-glass models in 
finite and infinite dimensions.

\appendix

\section{Parallel Tempering Ground-state Search}
\label{app:pt}

In this section we describe  the different numerical tools to
compute ground-state instances of the one-dimensional Ising chain fast.
As introduced in Refs.~\onlinecite{katzgraber:03} and \onlinecite{moreno:03}
we use parallel tempering Monte Carlo\cite{hukushima:96,marinari:96} to
calculate ground states. In Ref.~\onlinecite{katzgraber:03} it has already
been mentioned that parallel tempering Monte Carlo performs poorly for large
values of $\sigma$. In particular, for $\sigma \geq 2.5$ we find that
in practice the time
to find a ground state scales exponentially\cite{katzgraber:03f} in the system
size. In order to overcome this limitation we use the branch, 
cut, and price algorithm described below.

\begin{figure}
\includegraphics[width=\columnwidth]{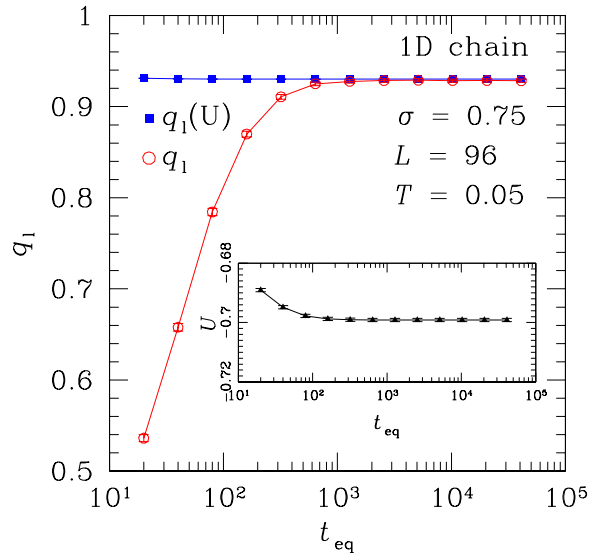}
\vspace*{-1.0cm}
\caption{(Color online)
Equilibration plot for the one-dimensional Ising chain: Average link overlap
as a function of Monte Carlo steps $t_{\rm eq}$ calculated directly
[Eq.~(\ref{eq:ql2})], and via the internal energy [Eq.~(\ref{eq:ql})] averaged
over the last half of the sweeps for $L = 96$, $T = 0.05$, and $\sigma =
0.75$. The data are equilibrated for $t_{\rm eq} \approx 10^4$ MCS, in the
simulations $4\times 10^4$ MCS have been used. Data for 2500 disorder
realizations.
}
\label{fig:equil}
\end{figure}

In the parallel tempering Monte Carlo method one simulates several identical
replicas of the system at different temperatures, and, in addition to the 
usual local moves, one performs global moves in which the temperatures 
of two replicas  with adjacent temperatures are exchanged. 
In this way, the temperature of a given replica wanders up and
down in a random manner, thus providing a more efficient sampling of the
energy landscape. For further details regarding the parallel tempering 
approach see Refs.~\onlinecite{hukushima:96} and \onlinecite{marinari:96}.
The parameters of the simulation are shown in Table \ref{tab:numbers}.  
If we take
the lowest temperature $T_{\rm min}$ to be $0.05$ ($T_{\rm min} \ll
T_{\rm c}$), then the minimum-energy state found at this temperature
is with very high probability the ground state. To test whether the
true ground state has been reached, four criteria have to be met: (i)
the same minimum-energy state has to be reached from two independent
replicas at $T_{\rm min}$ for all samples, and (ii) this state has to
be reached during $t_{\rm eq}$ sweeps in both copies. (iii) We
simulate for further $t_{\rm eq}$ sweeps to ensure that the energies
found do not change, and (iv) the system has to obey the equilibration
test for the one-dimensional Ising chain, introduced in
Ref.~\onlinecite{katzgraber:03}.  In this test the link overlap
$q_{\rm l}$ has to equate the link overlap calculated from the
internal energy $q_{\rm l}(U)$ via the relation
\begin{equation}
q_{\rm l} = 1 - \frac{2 T |[U]_{\rm av}/L|}{(T_c^{\rm MF})^2} \; ,
\label{eq:ql}
\end{equation}
where $T_{\rm c}^{\rm MF}$ is given by Eq.~(\ref{eq:tcmf}), $U$ is given by
Eq.~(\ref{eq:u}), and 
\begin{equation}
q_l = \frac{2}{N}\sum_{i,j} \frac{[J_{ij}^2]_{\rm av}}{(T_c^{MF})^2}
[ \langle S_i S_j \rangle^2 ]_{\rm av} \; .
\label{eq:ql2}
\end{equation}
Once both sides of Eq.~(\ref{eq:ql}) agree, the system is in equilibrium (see
Fig.~\ref{fig:equil}). Note that this is the case for the parameters listed in
Table \ref{tab:numbers}. If any of the aforementioned criteria are not met
(usually one instance in $10^5$), the calculated ground-state instance is 
rejected.

\begin{table}
\caption{
Parameters of the parallel tempering Monte Carlo simulations. 
The table shows the total 
number of Monte Carlo steps $t_{\rm eq}$ used for each value of $\sigma$ and 
$L$. We use between 10 and 17 temperatures, depending on
the system size, to ensure that the acceptance ratios of the parallel
tempering moves are larger than $\sim 0.30$. The lowest temperature used 
is $0.05$, the highest $1.70$. For the internal energy distributions 
(Sec.~\ref{1dchain:tgt0}) we compute
thermally averaged values of the internal energy for a given disorder
realization after equilibrating for $t_{\rm eq}$ Monte Carlo steps. The
averages are done over another period of
$t_{\rm eq}$ Monte Carlo steps. For $\sigma = 2.50$ and
$L = 96$ the calculations have been done using the BCP algorithm (Appendix
\ref{app:bcp}).
\label{tab:numbers}
}
\begin{tabular*}{\columnwidth}{@{\extracolsep{\fill}} c c c c c c }
\hline
\hline
$\sigma$ &  $8, 12, 16$  & $24, 32$ & $48, 64$ & $96, 128$ & $192$ \\
\hline
0.00 & $2\times 10^3$ & $4\times 10^3$ & $8\times 10^3$ & $4\times 10^4$ & $12\times 10^4$ \\
0.10 & $2\times 10^3$ & $4\times 10^3$ & $8\times 10^3$ & $4\times 10^4$ & $12\times 10^4$ \\
0.35 & $2\times 10^3$ & $4\times 10^3$ & $8\times 10^3$ & $4\times 10^4$ & $12\times 10^4$ \\
0.50 & $2\times 10^3$ & $4\times 10^3$ & $8\times 10^3$ & $4\times 10^4$ & $12\times 10^4$ \\
0.75 & $2\times 10^3$ & $4\times 10^3$ & $8\times 10^3$ & $4\times 10^4$ & $12\times 10^4$ \\
1.00 & $2\times 10^3$ & $4\times 10^3$ & $8\times 10^3$ & $8\times 10^4$ & $6\times 10^5$ \\
1.25 & $2\times 10^3$ & $4\times 10^3$ & $6\times 10^4$ & $6\times 10^5$ & \\
2.50 & $2\times 10^3$ & $4\times 10^3$ & $2\times 10^5$ & & \\
\hline
\hline
\end{tabular*}
\end{table}

\section{Branch, Cut, and Price Algorithm}
\label{app:bcp}

In this section we briefly explain how exact ground states of one-dimensional 
Ising spin-glass instances can be computed fast
for large values of
$\sigma$. To this end, we extend the branch-and-cut approach to a branch, 
cut, and price (BCP) method originating
in combinatorial optimization.
Since this approach has not yet been applied for spin
glasses, 
we give more details in the following section and 
discuss the performance in a subsequent section. Again, for
the fundamentals of the applied algorithm, i.e., the standard branch \& cut
approach, we refer the reader to Refs.~\onlinecite{barahona:88} and 
\onlinecite{liers:04}.

\subsection{Algorithm}

The problem of determining a ground state of an Ising spin-glass instance is
equivalent to determining a \emph{maximum cut} in the interaction
graph associated with the system.\cite{barahona:88} In the maximum cut 
problem we are given a graph $G=(V,E)$ with nodes $V$ 
and  edges $E$.
The nodes correspond to the spin sites, the edges to the
bonds. Weights $c_{ij}\in {\mathbb R}$ are given for all edges 
$ij\in E$. Let $W\subset V$ be a subset of nodes. The \emph{cut} $\delta(W)$ is
defined as the set of edges having exactly one endpoint in $W$. The
weight of a cut $\delta(W)$ is the sum of the weights of the
edges in the cut, and the maximum cut problem is to find a
cut $\delta(W)$ of $G$
with maximum weight among all possible node sets $W$. Determining a
ground state of a spin-glass instance amounts to calculating a maximum
cut in the interaction graph of the system, with edge weights chosen
as $c_{ij}=-J_{ij}$ for $ij\in E$.

The maximum cut problem is NP hard which makes it unlikely that there
exists a solution algorithm running in a number of steps bounded by a
polynomial in the size of the input. In practice, maximum cuts of
reasonably sized instances can be determined exactly by using the
branch-and-cut method from combinatorial optimization that has
exponential worst-case running time. For an instance, we always
maintain an upper and a lower bound for the optimum solution value of
the maximum cut. Iteratively we improve upper and lower bounds until
they are tight enough for proving optimality of a known solution. In
the upper bound computations, a sequence of \emph{linear programs} is
solved. Solving a linear problem amounts to optimizing a linear
objective function subject to a set of linear
constraints. Details are explained in
Refs.~\onlinecite{barahona:88} and \onlinecite{liers:04}.

For an instance of the one-dimensional Ising chain with $L=100$ spins
and $\sigma = 3.0$, the default version of the branch-and-cut
algorithm needs roughly 3 h CPU time on average on a 1400 MHz Athlon
processor. By extending the branch-and-cut algorithm to a
\emph{branch, cut, and price} algorithm we achieve a better performance.
Details about pricing algorithms can be found in Ref.~\onlinecite{juenger:95}.

The underlying idea of a pricing algorithm is as follows. There exists
a variable for each edge $ij\in E$, and we use the terms edge and
variable interchangeably. In the pure branch-and-cut algorithm we 
always work on the complete set of variables. However, in the extended
algorithm we start doing branch-and-cut, but only work on a small 
fraction of all variables. We add necessary variables (and delete
unnecessary ones) dynamically during the optimization process. This is
done in the so-called \emph{pricing} routine.

For the one-dimensional Ising chain, we make the assumption that for
big enough values of the parameter $\sigma$ the ``long-range couplings''
between two spins ``far apart'' from each other in the chain do not
strongly affect the ground state and can be neglected
temporarily. Thus, we start
working on a graph $G=(V,E)$ consisting of all nodes but only of a
fraction of all edges.  In our tests it performed best when the input
graph consisted of the $k\%$ edges with highest weights, measured in
absolute value, where the parameter $k$ is suitably chosen in order to
minimize the total running time. (For example, for $\sigma=3.0$ $k=20$ is a
good choice, for smaller $\sigma$ the value of $k$ is increased.) At
well-defined steps in the algorithm the pricing routine checks whether
there exists a (yet neglected) variable that has to be included in the
variable set for maintaining correctness. If no variable is added, and
upper and lower bounds are tight enough, we can prove optimality, and
stop. For our model, we can further improve the quality of the upper
bound within the BCP algorithm by separating not only
the cycle inequalities \cite{barahona:88} but also separating
heuristically the so-called parachute inequalities \cite{deza:97}
resulting in an improved bound and an additional speedup.

When the BCP algorithm is used, solving systems for
$\sigma=3.0$ takes on average $426 \pm 55$ seconds for $L=100$ on the
same 1400 MHz Athlon processor that needed three hours on average for
solving the same systems by branch-and-cut. In 
Ref.~\onlinecite{katzgraber:03} it is
reported that parallel tempering is less efficient in finding the
ground state for bigger values of $\sigma$, because parallel tempering needs 
longer to relax an inconvenient configuration. With the exact algorithm, in
contrast, we expect pricing to be only effective for bigger $\sigma$.
In this case we expect a speedup by using sparse graph techniques as
explained above.  For small $\sigma$ instead, the system is of the 
long-range type, and in the worst-case all neglected edges would have
to be added in the pricing routine.

In the following section we experimentally determine the running time
dependence of the BCP algorithm and its dependence
on the parameter $\sigma$ and on the system size. 

\subsection{Performance of the Algorithm}

In this section we study the performance of the BCP 
algorithm for the one-dimensional Ising chain model. We compute
ground states of samples for different system sizes $L$ and values of
the parameter $\sigma$. The studied ranges are 
$\sigma\in \{1.0,1.5,2.0,2.5,3.0\}$, 
and $L\leq 96$. We compute between 1000 and 6000 samples
per size and $\sigma$ value for small- and medium-sized instances and
at least 100 samples for the largest instances. All runs are
performed on a Linux cluster of identical AMD Athlon 1800+ machines.
Instances of size $L\leq 48$ and $\sigma \geq 2.0$ are solved within
seconds; for $\sigma=3.0$, computing a ground state of $L=280$ spins
takes on average $5161 \pm 275$ s. The hardest instances, $L=96,
\sigma=1.5$ needs up to a day computing time on one processor.

\begin{figure}
\includegraphics[width=\columnwidth]{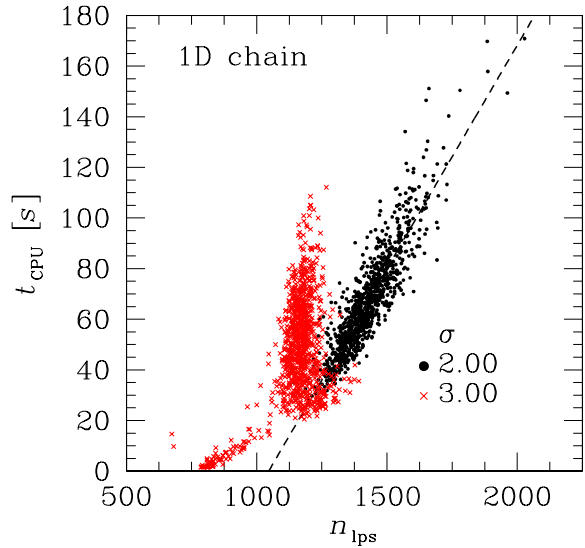}
\vspace*{-1.0cm}
\caption{(Color online)
Scatter plot for the CPU time $t_{\rm CPU}$ in seconds versus the number 
of lps $n_{\rm lps}$ for 1000 randomly chosen samples with $L=64$. 
The data for $\sigma = 2.0$ (black dots) are strongly correlated. The dashed
line is a guide to the eye. In contrast, data for $\sigma = 3.0$ (red crosses)
show strong sample-to-sample variations.
}
\label{fig:scatter_64_2}
\end{figure}

As argued before in Ref.~\onlinecite{liers:03}, there is no easy and ``ideal''
performance measure for a branch-and-cut algorithm. This remains true
for its extension to the BCP algorithm. As a measure of
the performance of the latter, we could use the needed CPU time which
however is machine dependent, or the number of solved linear programs
(lps), see Refs.~\onlinecite{palassini:03b} and \onlinecite{liers:03}. 
For $\sigma \lesssim 2$ we
find that the number of lps $n_{\rm lps}$ is strongly and almost linearly 
correlated with the CPU time $t_{\rm CPU}$, 
see Fig.~\ref{fig:scatter_64_2}. The same is
true for the pure branch-and-cut algorithm. However, for $\sigma
\gtrsim 2.0$ the CPU time for solving a lp considerably varies between
different samples of the same size, as can be seen for $L=64$ and
$\sigma=3.0$ in the scatter plot, Fig.~\ref{fig:scatter_64_2}.

A reason for this behavior is the following: In order to keep the
program flexible, in each iteration we both add new constraints to the
current linear program and remove constraints that once have been
added but have turned out to be unimportant. (Re-)optimizing a lp is
very fast if only a small number of constraints changes from one
iteration to the next but takes considerably longer if a substantial
change occurs. In the pricing extension, we start working on a subset
of the variables and might add further variables as explained above.
Possibly
a ``bad'' subset of variables is
chosen, in the sense that many of the added constraints become
unimportant later and are removed again. Then the lps change
considerably and their solution takes long.
This is more probable for big $\sigma$, as we start working
on a small subset of the variables. For smaller values of $\sigma$
instead, we start working on a bigger fraction of all variables and
find a stronger correlation between number of lps and CPU
time.\cite{bcp} Given the broad variation in the CPU time per lp for
some values of $\sigma$, we use the mean of the CPU time as a
performance measure. We notice that the figures remain qualitatively
comparable when the mean of the linear programs is taken instead of
the CPU time. We have checked that the mean of both the CPU time and
the number of linear programs is defined for our sampling as the
distribution shows a pronounced tail. Performing a detailed
statistical analysis we show that the data are thin-tail
distributed\cite{gumbel:58} with a well-defined mean.

\begin{figure}
\includegraphics[width=\columnwidth]{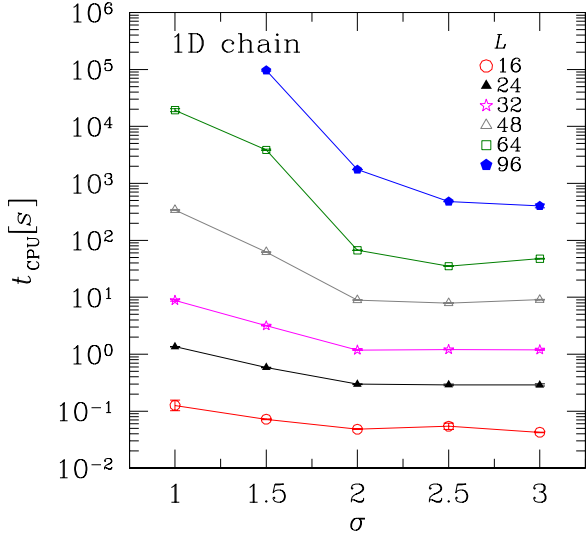}
\vspace*{-1.0cm}
\caption{(Color online)
Mean CPU time $t_{\rm CPU}$ in seconds for determining a ground state 
versus $\sigma$ for different $L$ in a linear-log scale. For increasing
$\sigma$ and for all system sizes $L$ the time to find a ground-state
configuration decreases thus showing that the algorithm becomes more 
efficient when the interactions are more short-ranged ($\sigma \rightarrow
\infty$).
}
\label{fig:cpuSigma}
\end{figure}

In Fig.~\ref{fig:cpuSigma} we show the average CPU time for solving
an instance as a function of $\sigma$, for different system sizes $L$.
Ground states are computed fast for big values of the
parameter $\sigma$, whereas it takes considerably longer for smaller
$\sigma\leq 1.5$. This effect becomes more apparent with increasing 
system size $L$.

\begin{figure}
\includegraphics[width=\columnwidth]{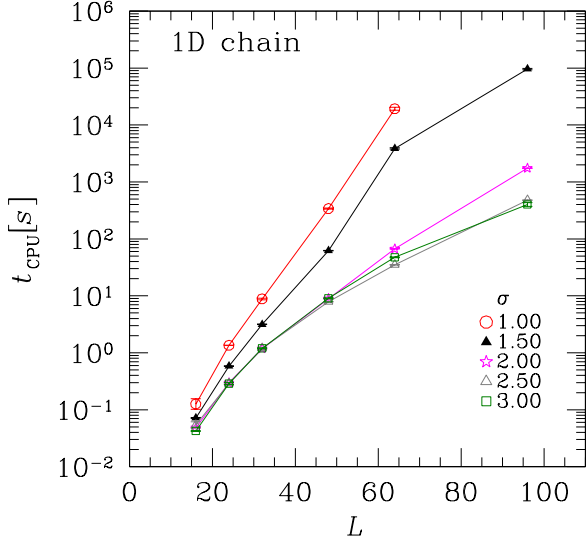}
\vspace*{-1.0cm}
\caption{(Color online)
Mean CPU time for determining a ground state versus the
system size $L$ for different values of $\sigma$ in a linear-log
scale. Note that the CPU time increases slower than exponential for all 
values of $\sigma$ studied. For $\sigma \ge 2.5$ the CPU time increases with
a power of the system size.
}
\label{fig:cpuL}
\end{figure}

We also study the CPU time as a function of the
total number of edges, i.e., the total number of variables, for
different values of $\sigma$. The increase in the CPU time with the
number of variables is consistent with a polynomial dependency, even
for the smallest studied value of $\sigma$. When fitting a function of
the form $f(m)\sim am^b$, with $m$ being the number of variables (bonds),
we obtain $a=0.009\pm 0.007$, $b = 1.3\pm 0.1$ for
$\sigma = 2.0$. A similar behavior can be found when studying the CPU time as
a function of system size $L$, see Fig.~\ref{fig:cpuL}.

A qualitatively similar behavior can be found in the data when plotted 
as a function of lps instead of CPU time (not shown).

\begin{acknowledgments}
We would like to thank I.~A.~Campbell, P.~C.~Holdsworth, O.~Martin, 
S.~Trebst, M.~Troyer, D.~W\"urtz, and A.~P.~Young for discussions. 
In particular we thank A.~P.~Young for spotting a mistake in the manuscript.
A.K.H.~obtained financial support from the {\em VolkswagenStiftung} 
(Germany) within the program ``Nachwuchsgruppen an Universit\"aten.''
F.L.~has been supported by the German Science Foundation
(DFG) in Project No.~Ju 204/9-1. The calculations were performed 
on the Asgard cluster at ETH Z\"urich and on the SCALE
cluster of E.~Speckenmeyer's group in Cologne. We are indebted to 
M.~Troyer and G.~Sigut for allowing us to use the idle time on the 
Asgard cluster.
\end{acknowledgments}

\bibliography{refs,comment}

\begin{thebibliography}{60}
\bibitem[{dis()}]{distances}
\bibinfo{note}{Note that the distance $r_{ij}$ between two spins $i$ and $j$ on
  the chain is determined by $r_{ij} = 2R\sin(\alpha/2)$, where $R$ is the
  radius of the chain and $\alpha$ is the angle between the two sites on the
  circle. The previous expression can be rewritten in terms of the system size
  $L$ and the positions of the spins to obtain $r_{ij} = (L/\pi)\sin\left(\pi
  |i - j|/L\right)$.}
\bibitem[{sti()}]{stiffness}
\bibinfo{note}{The stiffness exponent $\theta$ describes a measure of the
  typical free energy difference between two coupling-independent boundary
  conditions. Usually it is defined via the change in energy after inducing a
  domain wall trough a change in boundary conditions, i.e., $\Delta E =
  [|E_{\rm p} - E_{\rm ap}|]_{\rm av} \sim L^\theta$ with $E_{\rm p}$ the
  ground-state energy of the system with periodic, and $E_{\rm ap}$ the
  ground-state energy of the system with antiperiodic boundary conditions. When
  $\theta < 0$ the system is unstable toward excitations, therefore $T_{\rm
  c}=0$.}
\bibitem[{bcp()}]{bcp}
\bibinfo{note}{Another possibility for the broad distribution of the running
  time per lp could be a strong variation in the number of added and deleted
  variables. However, in our sampling this was not the case.}
\end{thebibliography}

\end{document}